\documentclass[11pt]{article}
\usepackage{amsfonts,amssymb,amsmath}
\DeclareMathAlphabet{\mathrsfs}{U}{rsfs}{m}{n}

\def \x {{\mathrm{x}}}
\def \y {{\mathrm{y}}}

\def \z {{\mathrm{z}}}

\def \R {{\mathbb R}}
\def \C {{\mathbb C}}
\def \H {{\mathbb H}}
\def \F {{\mathbb F}}

\def \LL {\mathcal{L}}
\def \MA {\mathcal{M}}

\def \spr {\cdot}
\def \eps {\varepsilon}

\def \Young {\mathrsfs{Y}}
\def \LW {:{\hspace{-2pt}}}
\def \RW {{\hspace{-2pt}}:}
\def \vac {|\hspace{-1pt}\text{\textit{vac}}\hspace{-1pt}\rangle}

\renewcommand{\geq}{\geqslant}
\renewcommand{\leq}{\leqslant}

\newcommand{\VEC}[1]{\vec{#1}}
\newcommand{\trp}[1]{\hspace{0.5pt}{^t}\hspace{-1pt}{#1}\hspace{0.5pt}}

\newcommand{\mbf}[1]{\ensuremath{\mathchoice
                    {\mbox{\boldmath$\displaystyle\mathbf{\mathit{#1}}$}}
                    {\mbox{\boldmath$\textstyle\mathbf{\mathit{#1}}$}}
                    {\mbox{\boldmath$\scriptstyle\mathbf{\mathit{#1}}$}}
                    {\mbox{\boldmath$\scriptscriptstyle\mathbf{\mathit{#1}}$}}}}
\newcommand{\Mbf}[1]{\ensuremath{\mathchoice
                    {\mbox{\boldmath$\displaystyle\mathbf{#1}$}}
                    {\mbox{\boldmath$\textstyle\mathbf{#1}$}}
                    {\mbox{\boldmath$\scriptstyle\mathbf{#1}$}}
                    {\mbox{\boldmath$\scriptscriptstyle\mathbf{#1}$}}}}

\newcommand{\beq}{\begin{equation}}
\newcommand{\eeq}{\end{equation}}
\newcommand{\beqa}{\begin{eqnarray}}
\newcommand{\eeqa}{\end{eqnarray}}
\newcommand{\nn}{\nonumber \\}
\def \podr {&& \hspace{-15pt}}

\newcounter{Theorem}

\newcounter{Definition}

\renewcommand{\theequation}{\arabic{section}.\arabic{equation}}
\def \setcntrs {\setcounter{equation}{0}\setcounter{Theorem}{0}\setcounter{Definition}{0}}

\newcounter{tmpc}
\newlength{\tmplenght}
\newlength{\tmplenghta}
\newlength{\tmplenghtb}
\newlength{\tmplenghtc}

\newenvironment{LIST}[1]{%
\setlength{\tmplenghta}{#1}
\setlength{\tmplenghtb}{#1}
\setlength{\tmplenghtc}{#1}
\advance\tmplenghtb-5pt
\advance\tmplenghtc 42pt
\setcounter{tmpc}{0}
\begin{list}{{\rm (\alph{tmpc})}}{\usecounter{tmpc}
\setlength{\leftmargin}{\tmplenghta}
\setlength{\rightmargin}{0cm}
\setlength{\itemsep}{1pt}
\setlength{\topsep}{3pt}
\setlength{\labelsep}{5pt}
\setlength{\labelwidth}{\tmplenghtb}
\setlength{\listparindent}{\tmplenghta}}
}{\end{list}}

\title{{\bf Infinite dimensional Lie algebras in 4D conformal quantum
  field theory}%
\footnote{Lecture at the Workshops ``Lie Theory and Its Applications in
Physics'', 18--24 June 2007, Varna, Bulgaria; ``Infinite Dimensional
Algebras and Quantum Integrable Systems'', 23--27 July, 2007, Faro, Portugal;
and ``Supersymmetries and Quantum Symmetries'', 30 July--4 August, 2007,
Dubna, Russia}}

\author{Bojko Bakalov$^{1}$, \\[2mm]
Nikolay M.\ Nikolov$^{2}$, \\[2mm]
Karl-Henning Rehren$^{2,3}$, \\[2mm]
Ivan Todorov$^{2}$}

\begin{document}

\maketitle

\begin{center}
\scriptsize
\vskip-5mm
\parbox{300pt}{
\begin{LIST}{21pt}
\item[$^{1}$]
Department of Mathematics, North Carolina State University, \\
Box 8205, Raleigh, NC 27695, USA; \\
bojko\_bakalov@ncsu.edu
\item[$^{2}$]
Institute for Nuclear Research and Nuclear Energy, \\
Tsarigradsko Chaussee 72, BG-1784 Sofia, Bulgaria; \\
mitov@inrne.bas.bg, todorov@inrne.bas.bg
\item[$^{3}$]
Institut f\"ur Theoretische Physik, Universit\"at G\"ottingen, \\
Friedrich-Hund-Platz 1, D-37077 G\"ottingen, Germany; \\
rehren@theorie.physik.uni-goe.de
\end{LIST}}
\end{center}

\begin{abstract}
\noindent
The concept of global conformal invariance (GCI) opens the way of
applying algebraic techniques, developed in the context of 2-dimensional chiral
conformal field theory, to a higher (even) dimensional space-time. In
particular, a system of GCI scalar fields of conformal dimension two gives
rise to a Lie algebra of harmonic bilocal fields, $V_M(x,y)$, where
the $M$ span a finite dimensional real matrix algebra $\MA$ closed under
transposition. The associative algebra $\MA$ is irreducible iff its commutant
$\MA'$ coincides with one of the three real division rings. The Lie algebra
of (the modes of) the bilocal fields is in each case an infinite dimensional
Lie algebra: a central extension of $sp(\infty, \mathbb R)$
corresponding to the field ${\mathbb R}$ of reals, of
$u(\infty, \infty)$ associated to the field ${\mathbb C}$ of complex numbers,
and of $so^*(4\infty)$ related to the algebra ${\mathbb H}$ of quaternions.
They give rise to quantum field theory models with superselection sectors
governed by the (global) gauge groups $O(N), U(N)$, and
$U(N,{\mathbb H})=Sp(2N)$, respectively.
\end{abstract}


\section{Introduction}\label{sec1}
\setcntrs

The assumption of global conformal invariance -- which says that we are
dealing with a single valued  representation of $SU(2,2)$ rather than
with a representation of its covering -- in  (4-dimensional) Minkowski
space has surprisingly strong consequences \cite{NT01}. Combined with
the Wightman axioms, it implies \emph{Huygens locality}, which yields
the vertex-algebra-type condition
\beq\label{eq1.1}
\bigl((\x-\y)^2\bigr)^n \, \bigl[\phi(\x), \psi(\y)\bigr] = 0
\quad \text{for} \quad n \gg 0 \eeq for any pair $\phi, \psi$ of local
Bose fields ($n\gg 0$ meaning ``$n$ sufficiently large''). Huygens
locality and energy positivity imply, in turn, rationality of
correlation functions. A GCI quantum field theory (QFT) that admits a
stress-energy tensor (something, we here assume) necessarily involves
infinitely many conserved symmetric tensor currents in the operator
product expansion (OPE) of any Wightman field with its conjugate.
The twist two contributions give rise to a harmonic {\it bifield}
$V(\x,\y)$, which is an important tool in the study of GCI QFT models
\cite{NST02,NST03,NRT05,BNRT07,NRT07}.  The spectacular development of
2-dimensional (2D) \emph{conformal field theory} in the 1980's
is based on the preceding study of infinite dimensional (Kac--Moody
and Virasoro) Lie algebras and their representations. A
straightforward generalization of this tool did not seem to apply
in higher dimensions. After the first attempts
to construct (4D) Poincar{\'e} invariant Lie fields led to examples
violating energy positivity \cite{L67}, it was proven \cite{B76},
that scalar Lie fields do not exist in three or more dimensions.
It is therefore important to realize that the argument does not pass to
{\em bi\/}fields, and that the above mentioned harmonic bifields do
give rise to infinite dimensional Lie algebras.

Consider bilocal fields of the form
\beq \label{eq1.1a}
V_M(\x,\y) \, = \, \sum_{ij} \, M_{ij} \ {:}\varphi_i(\x)\varphi_j(\y){:}
\,, \eeq
where $M$ is a real matrix and $\varphi_j$ are a system of independent
real massless free fields. According to Wick's theorem, the
commutator of $V_{M_1}(\x_1,\x_2)$ and $V_{M_2}(\x_3,\x_4)$ is:
\beqa
\label{eq2.1n}
&& \hspace{-20pt}
\bigl[V_{M_1}(\x_1,\x_2), V_{M_2}(\x_3,\x_4)\bigr]
\, = \,
\Delta_{2,3} \, V_{M_1M_2}(\x_1,\x_4)
\, + \,
\Delta_{2,4} \, V_{M_1\trp{M_2}}(\x_1,\x_3)
\nn && \hspace{-20pt}
\qquad \hspace{10pt}
\, + \
\Delta_{1,3} \, V_{\trp{M_1}M_2}(\x_2,\x_4)
\, + \,
\Delta_{1,4} \, V_{\trp{M_1}\trp{M_2}}(\x_2,\x_3)
\nn && \hspace{-20pt}
\qquad \hspace{10pt}
\, + \
\mathrm{tr} \, (M_1 M_2) \Delta_{12,34} + \mathrm{tr} \, (\trp{M_1}
M_2) \Delta_{12,43}, 
\eeqa
where $\trp{M}$ is the transposed matrix, $\Delta_{j,k}$ is the free field
commutator, $\Delta_{j,k}$ $=$ $\Delta^+_{j,k} - \Delta^+_{k,j}$,\ and
$\Delta_{jk,lm}$ $=$ $\Delta^+_{j,m}\Delta^+_{k,l} - \Delta^+_{m,j}\Delta^+_{l,k}$\
for\ $\Delta^+_{j,k}$ $:=$ $\Delta_+(\x_j$ $-$ $\x_k)$, the two point
massless scalar correlation function.

It is one of the main results of \cite{NRT07} that the same abstract
structure can be derived from first principles in GCI quantum field theory.
More precisely, the twist two bilocal fields appearing in the OPE of
any two scalar fields of dimension 2 can be linearly labeled by
matrices $M$ such that the commutation relations \eqref{eq2.1n} hold. {}
From this, the representation \eqref{eq1.1a} can be deduced.
In the present paper we shall consider only finite size matrices; in
general, the system of independent massless free fields can be infinite
and then the $M$'s should be assumed to be Hilbert--Schmidt operators.

The question arises, whether there are nontrivial linear subspaces
$\MA$ of real matrix algebras upon which the commutation
relations of the corresponding bifields $V_M$ ($M \in \MA$) close.
We shall call such systems of bifields \textit{Lie systems},
or, \textit{Lie bifields}.
It follows from (\ref{eq2.1n}) that if $\MA$ is a {\em $t$-subalgebra}
(i.e., a subalgebra closed under transposition)
of the real matrix algebra, then $\{V_M\}_{M \, \in \, \MA}$
is a Lie system. Conversely, any Lie system corresponds to a subalgebra
$\MA$ such that $\trp{M_1}M_2$, $M_1\trp{M_2}$,
$\trp{M_1}\trp{M_2}$ $\in\MA$ whenever $M_1,M_2 \in\MA$.
In particular, if $\MA$ contains the identity matrix,
then it is a $t$-subalgebra.

\section{$t$-subalgebras of real matrix algebras}\label{sec2a}
\setcntrs

Let us consider $t$-subalgebras $\MA$ of the matrix algebra $Mat(L,\R)$,
where $L$
is a positive integer (equal to the number of fields $\varphi_j$).
The classification of all such $\MA$ is a classical mathematical problem,
which goes back to F.G.~Frobenius, I.~Schur, and J.H.M.~Wedderburn
(see, e.g., \cite[Chapter XVII]{L02} and \cite[Chapter 9, Appendix II]{B82}).

We first observe that $\MA$ is equipped with the Frobenius inner product
\beq\label{eq2.1xx}
\langle M_1, M_2 \rangle = \mathrm{tr} \, (\trp{M_1}M_2) = \sum_{ij} \,
(M_1)_{ij} \, (M_2)_{ij} \,,
\eeq%
which is symmetric, positive definite, and has the property
$\langle M_1 M_2,\,M_3 \rangle=\langle M_1,$ $M_3\;\trp{M_2} \rangle$.
This implies that for every right ideal $\mathcal{I} \subset\MA$,
the orthogonal complement $\mathcal{I}^\perp$ is again a right ideal.
Note also that $\mathcal{I}$ is a right ideal if and only if
$\trp{\mathcal{I}}$ is a left ideal.
Therefore, $\MA$ is a {\em semisimple} algebra (i.e., a direct sum of
left ideals), and every module over $\MA$ is a direct sum of
irreducible ones.

Now assume, without loss of generality, that the algebra
$\MA \subset End_{\,\R}\ \LL \cong Mat(L,\R)$
acts irreducibly on the vector space $\LL \cong \R^L$.
Let $\MA'\subset End_{\,\R}\ \LL$ be the {\em commutant} of $\MA$,
i.e., the set of all matrices $M$ commuting with all elements of $\MA$.
Then by Schur's lemma
(whose real version \cite{L02} is much less popular than the complex one),
$\MA'$ is a real division algebra.
By the Frobenius theorem, $\MA'$ is isomorphic to $\R$, $\C$, or $\H$
as a real algebra (where $\H$ denotes the algebra of {\em quaternions}).
Finally, the classical Wedderburn theorem gives that
$\MA$ is isomorphic to the matrix algebra $End_{\MA'}\LL$.
In addition, since $\MA$ is closed under transposition, then $\MA'$
is also a $t$-algebra, and the transposition in $\MA'$
coincides with the conjugation in $\R$, $\C$, or $\H$, respectively.

Observe that, since $\MA \cong End_\F\,\LL$
(where $\F=\R$, $\C$, or $\H$),
we can view $\LL$ as a left $\F$-module on which $\MA$ acts
$\F$-linearly. Alternatively, $\LL$ can be made an $(\MA,\F)$-bimodule
by setting $M \cdot f \cdot M' := M \, (\trp{M'}) \, f$ for
$f \in \LL$, $M \in \MA$ and $M' \in \MA' \cong \F$.
Then the embedding $\F\subset End_\F\,\LL\cong\MA$ endows $\LL$
with the structure of an $\F$-bimodule. In other words,
we have two commuting copies,
left and right, of $\F$ in $End_{\,\R}\ \LL$,
which are subalgebras of $\MA$ and $\MA'$, respectively.
Moreover, denoting $N=dim_{\F} \, \LL$, we have:
$L = dim_{\R} \, \LL = N \, dim_{\R} \, \F =N,2N$ or $4N$
when $\F=\R,\C$ or $\H$, respectively.

If $\MA$ is not an irreducible $t$-subalgebra of $Mat(L,\R)$,
i.e., $\LL \cong \R^L$ is not an irreducible $\MA$-module,
then $\LL$ splits into irreducible submodules, each of them of the above
three types:
\beq\label{e1.3}
\LL \, = \, \LL_{\R} \, \oplus \, \LL_{\C} \, \oplus \, \LL_{\H} \,,
\eeq where each $\LL_{\F}$ ($\F=\R,\C,\H$)
is an $\F$-module such that $\MA$ acts on it $\F$-linearly.
 In our QFT application, the space $\LL$ is the real linear span of the
real massless scalar fields $\varphi_j$, and then a Lie system of bifields
$V_{M}$ splits into three subsystems: of types $\R$, $\C$ and $\H$.
The first two cases were considered in a previous paper \cite{BNRT07}
and led to gauge groups of type $U(N,\R) = O(N)$ and $U(N,\C) = U(N)$, respectively,
where $N=dim_{\F} \, \LL$.
Here we are going to consider the third case in which, as we shall see,
the gauge groups that arise are of type $U(N,\H) = Sp(2N)$,
the compact real form of the {\em symplectic group}.

In each of the three cases, the associated infinite-dimensional
Lie algebra (\ref{eq2.1n}) has a central charge proportional to the
order $N$ of the gauge group $U(N,\F)$.

\section{Irreducible Lie bifields and associated dual pairs}\label{sec2}
\setcntrs

In this section we consider Lie bifields $\{V_{M}\}_{M\in\MA}$
corresponding to irreducible $t$-subalgebras $\MA$ of $Mat(L,\R)$.
As discussed in the previous section, we have
$\MA\cong End_{\MA'}\LL$, where $\LL \cong \R^L$ and
the commutant $\MA'\cong$ $\R$, $\C$, or~$\H$.

In the case when $\MA'\cong\R$ and $dim_\R\, \LL=1$, we have one bifield
\beq
\label{eq2.1a}
V(\x,\y) \, = \, {:}\varphi(\x) \varphi(\y){:} \,. \eeq
More generally, $V$ can be taken a sum of $N$ independent copies
of Lie bifields of type (\ref{eq2.1a}),
\beq
\label{eq2.2a}
V(\x,\y) \equiv V_{(N)}(\x,\y) \, = {:}\mbf{\varphi}(\x)
\mbf{\varphi}(\y){:} \, =  \mathop{\sum}\limits_{j \, = \, 1}^N
{:}\varphi_j(\x) \varphi_j(\y){:} \,,
\eeq
which is invariant under the \textit{gauge group} $O(N)$ (including
reflections). Here $L=N$ and $O(N)$ is realized as
the group of linear automorphisms of $\LL =
Span_{\R} \{\varphi_j\}$ preserving the quadratic form
(\ref{eq2.2a}) in $\varphi_j$. In this case the
\textit{field Lie algebra} (i.e., the Lie algebra of field modes
corresponding to the eigenvalues of the one-particle energy, see the Appendix)
is isomorphic to a central extension of $sp(\infty,\R)$ of central charge $N$; see \cite{BNRT07}.

The case when $\MA' \cong \C$ and $dim_\C\, \LL=1$ is given by two
real bifields, $V_{\Mbf{1}}$ and $V_{\eps}$ that correspond to the $2
\times 2$ matrices
\beq\label{eq2.3a}
\Mbf{1} \, = \, \left(\hspace{-4pt} \begin{array}{cc} 1 & 0 \\ 0 & 1 \end{array} \hspace{-2pt}\right)
\, , \quad
\varepsilon \, = \, \left(\hspace{-4pt} \begin{array}{rc} 0 & 1 \\ -1 & 0 \end{array} \hspace{-2pt}\right) .
\eeq
They are thus generated by two independent real massless fields $\varphi_1 (\x)$ and $\varphi_2(\x)$:
\beqa\label{eq2.4a}
V_{\Mbf{1}} (\x,\y) \, = \podr {:}\varphi_1(\x) \varphi_1(\y){:} + {:}\varphi_2(\x) \varphi_2(\y){:} \,,
\nn
V_{\varepsilon} (\x,\y) \, = \podr {:}\varphi_1(\x) \varphi_2(\y){:} - {:}\varphi_2(\x) \varphi_1(\y){:}
\,.
\eeqa
Combining $\varphi_1$ and $\varphi_2$ into one complex field
$\mbf{\varphi} (\x) = \varphi_1 (\x) + i \varphi_2(\x)$ we get that
$V_{\Mbf{1}}$ and $V_\eps$ are the real and the imaginary parts of the
complex bifield
\beq\label{eq2.5a}
W (\x,\y) \, = \, {:}\mbf{\varphi}^* (\x) \mbf{\varphi} (\y){:}
\, = \, V_{\Mbf{1}} (\x,\y) + i \, V_\eps (\x,\y) .
\eeq
Taking again $N$ independent copies of such Lie bifields,
\beq\label{eq2.6a}
W_{(N)} (\x,\y) \, = \, \mathop{\sum}\limits_{j \, = \, 1}^N  {:}\mbf{\varphi}^*_j (\x) \mbf{\varphi}_j (\y){:}
\, , \quad
\mbf{\varphi}_j (\x) \, = \, \varphi_{1,j} (\x) + i \, \varphi_{2,j} (\x),
\eeq
we get a {\em gauge group} $U (N)$, where $L=2N$.
The {\em field Lie algebra} in this second case is
isomorphic to a central extension of $u(\infty,\infty)$ again of central charge $N$ (\cite{BNRT07}).

Finally, for $\MA' = {\mathbb H}$ the minimal size of the matrices in
$\MA$ is four. We can formally derive the basic bifields $V_{M}$ in
this case as in the above complex case~(\ref{eq2.5a}). Let us combine
the four independent scalar fields $\varphi_j (\x)$ ($j=0,1,2,3$)
in a single ``quaternionic-valued'' field and its conjugate:
\beqa
\label{eq2.6}
\mbf{\varphi} (\x) \, = \podr \varphi_0 (\x) + \varphi_1 (\x) \, I + \varphi_2 (\x) \, J + \varphi_3 (\x) \, K
, \quad
\nn
\mbf{\varphi}^+ (\x) \, = \podr \varphi_0 (\x) - \varphi_1 (\x) \, I - \varphi_2 (\x) \, J - \varphi_3 (\x) \, K,
\eeqa
where $I, J, K$ are the (imaginary) quaternionic units
satisfying $IJ = K = -JI,
I^2 = J^2 = K^2 = -1$. This allows us to write a quaternionic bifield $Y$ as
\beq
\label{eq2.7}
Y(\x,\y) \, = \, {:}\mbf{\varphi}^+(\x) \, \mbf{\varphi}(\y){:}
\, = \,  V_0(\x,\y) + V_1(\x,\y) \, I + V_2(\x,\y) \, J + V_3(\x,\y) \, K,
\eeq
where the components $V_\alpha$ ($\alpha = 0, 1, 2, 3$) of $Y$ can be
further expressed in terms of the 4-vectors $\mbf{\varphi}$ and a
$4\times 4$ matrix realization of the quaternionic units in a manner
similar to~(\ref{eq2.4a}):
\beqa
\label{eq2.8}
&
V_\alpha(\x,\y) \, \equiv \, V_{\ell_{\alpha}} \bigl(\x,\y\bigr) \,
\, = \, {:}\mbf{\varphi}(x) \,\ell_\alpha \, \Mbf{\varphi}(y){:} \, , \quad
&
\nonumber \\
&
\nonumber \\
&
\ell_0 = \Mbf{1} ,\qquad
\ell_1 \, = \, \left(\hspace{0pt} \begin{array}{rrrr}                                                                                                                                       0 & 1 & 0 & 0 \\
{\hspace{-8pt}}-{\hspace{-2pt}}1 & 0 & 0 & 0 \\
0 & 0 & 0 & {\hspace{-8pt}}-{\hspace{-2pt}}1 \\
0 & 0 & 1 & 0
\end{array}
\right)
,\qquad
&
\nonumber \\
&
\ell_2 \, = \,
\left(\hspace{0pt}
\begin{array}{rrrr}
0 & 0 & 1 & 0 \\
0 & 0 & 0 & 1 \\
{\hspace{-8pt}}-{\hspace{-2pt}}1 & 0 & 0 & 0 \\
0 &{\hspace{-8pt}}-{\hspace{-2pt}}1 & 0 & 0
\end{array} \right)
,\qquad
\ell_3 \, = \,
\left(\hspace{0pt}
\begin{array}{rrrr}
0 & 0 & 0 & 1 \\
0 & 0 &{\hspace{-8pt}}-{\hspace{-2pt}}1 & 0 \\
0 & 1 & 0 & 0 \\
{\hspace{-8pt}}-{\hspace{-2pt}}1 & 0 & 0 & 0
\end{array} \right) .
\eeqa
It is straightforward to check that the $4\times 4$ matrices $\ell_{\alpha}$
generate the quaternionic algebra $\H \cong \MA$.
The commutant $\MA'$ in $Mat(4,\R)$ is spanned by the unit matrix and
another realization of the imaginary quaternionic units as a set of real
antisymmetric $4\times 4$ matrices $r_k$ $(k=1,2,3)$.
The two sets $\{r_k\}_{k\, = \, 1}^3$ and $\{\ell_k\}_{k\, = \, 1}^3$
correspond to the splitting of the Lie algebra $so(4)$
into a direct sum of two $so(3)$ algebras:
\beqa
\label{eq2.9}
&
\ell_1 \, = \, \sigma_3 \otimes \varepsilon
\, , \quad
\ell_2 \, = \, \varepsilon \otimes \Mbf{1}
\, , \quad
\ell_3 \, = \, \ell_1 \, \ell_2 \, = \, \sigma_1 \otimes \varepsilon \,,
\quad
&
\nn
&
r_1 \, = \, \varepsilon \otimes \sigma_3
\, , \quad
r_2 \, = \, \Mbf{1} \otimes \varepsilon
\, , \quad
r_3 = r_1 \, r_2 \, = -r_2 \,r_1 \, = \, \varepsilon \otimes \sigma_1  \,,
\quad
&
\eeqa
where $\sigma_k$
are the Pauli matrices and $\varepsilon = i\sigma_2$
as in (\ref{eq2.3a}).

We shall demonstrate that the quaternionic field $Y$ (\ref{eq2.7}) generates
a central extension of
the Lie algebra\footnote{%
For a description of the Lie algebra $so^* (2n)$ of the noncompact
group $SO^*(2n)$ 
and of its highest weight representations, see \cite{EHW83}. For an oscillator realization of the
Lie superalgebra $osp(2m^*\vert 2n)$ (with even subalgebra $so^*(2m)\times
sp(2n)$), see \cite{GS91}.
If we view $so^* (4\infty)$ as an inductive limit of $so^*(4n)$
then the central extension is trivial.}
$so^* (4\infty)$.
To this end, we represent $Y$ by a pair of complex bifields
\beq\label{eq3.11it}
\begin{array}{l}
W (\x,\y) = \, \frac{1}{2} \, \Bigl(V_0 (\x,\y) + i V_3
(\x,\y)\Bigr) \,
\\
\qquad \qquad \qquad
= \, \LW \psi^*_1 (\x) \, \psi_1 (\y) \RW
+ \LW \psi^*_2 (\x) \, \psi_2 (\y) \RW
\, = \, W (\y,\x)^*
\, , \quad
\\
\hfill
\\
A (\x,\y) = \, \frac{1}{2} \, \Bigl(V_1 (\x,\y) - i V_2(\x,\y)\Bigr)
\\
\qquad \qquad \qquad
= \, \psi_1 (\x) \, \psi_2 (\y) - \psi_2 (\x) \,
\psi_1 (\y) \, = \, - A (\y,\x)
\, , \qquad
\end{array}
\eeq
and their conjugates, where $\psi_{\alpha}$ are complex linear
combinations of $\varphi_{\nu}$:
\beq\label{eq3.12it}
\psi_1 \, = \, \frac1{\sqrt{2}} \Bigl( \varphi_0 + i \varphi_3 \Bigr)
\, , \quad
\psi_2 \, = \, \frac1{\sqrt{2}} \Bigl( \varphi_1 - i \varphi_2 \Bigr)\,.
\eeq

Substituting as above each $\varphi_{\nu}$ (respectively $\psi_{\alpha}$) by
an $N$-vector of commuting free fields we can write the nontrivial local
commutation relations (CR) of $W(1,2)$ $\equiv W (\x_1,\x_2)$ and $A(1,2)$
in the form
\beqa\label{eq3.13it}
\bigl[ W^* (1,2), W(3,4) \bigr]
\, = \podr
\Delta_{1,3} \, W(2,4) + \Delta_{2,4} \, W^* (1,3) + 2N \Delta_{12,43} \, ; \quad
\\ \label{eq3.14it}
\bigl[ W(1,2), A(3,4) \bigr]
\, = \podr
\Delta_{1,3} \, A(2,4) - \Delta_{1,4} \, A (2,3) \, , \quad
\nn
\bigl[ W(1,2), A^*(3,4) \bigr]
\, = \podr
\Delta_{2,3} \, A^* (1,4) - \Delta_{2,4} \, A^* (1,3) \, , \quad
\nn
\bigl[ A^*(1,2), A(3,4) \bigr]
\, = \podr
\Delta_{1,3} \, W(2,4) - \Delta_{1,4} \, W (2,3)
+\Delta_{2,4} \, W(1,3)
\nn \podr
- \, \Delta_{2,3} \, W (1,4) + 2N \bigl(\Delta_{12,43}-\Delta_{12,34}\bigr)
\,. \quad
\eeqa
In particular, $W$
coincides with $W_{(2N)}$ in (\ref{eq2.6a}) and
generates the $u (\infty,\infty)$ algebra (of even central charge),
which contains the compact Cartan subalgebra of $so^*
(4\infty)$; see Appendix~A. On the other hand, it is
straightforward to display the gauge group in the original
picture as the invariance group of the quaternionic valued bifield
$Y$~(\ref{eq2.7}) viewed as a quaternionic form in the
$N$-dimensional space of real quaternions. We obtain the group
of $N\times N$ unitary matrices with quaternionic entries
\beq\label{eq3.15it}
U (N,\H) \, = \, Sp(2N) \,\equiv\, USp(2N) \,,
\eeq
i.e., the compact group of unitary complex
symplectic $2N \times 2N$ matrices.

\section{Unitary positive energy representations and
superselection  structure}\label{sec4}

\setcntrs

Two important developments, one in QFT, the other in representation theory,
originated half a century ago from the talks of Rudolf Haag and Irving
Segal at the first Lille conference \cite{Lille57} on mathematical
problems in QFT. Later they gradually drifted apart and lost sight of
each other. The work of the Hamburg--Rome--G\"ottingen school on the
operator algebra approach to local quantum physics \cite{H92}
culminated in the theory of (global) gauge groups and superselection
sectors \cite{DR90,BDLR92}. The parallel development of the theory of
highest weight modules of semisimple Lie groups (and of the related
dual pairs) can be traced back from \cite{EHW83, H89, S90}. Here we aim at
completing the task, undertaken in \cite{BNRT07} of (restoring and)
displaying the relationship between the two developments.

Before formulating the main result of this section we shall rewrite
the CR (\ref{eq3.13it}), (\ref{eq3.14it}) in terms of the discrete
modes of $W, A$ and $A^*$ and introduce along the way the conformal
Hamiltonian. We first list the $u(\infty, \infty)$ modes of $W$
\cite{BNRT07} and write down their CR. Here belong the generators
$E_{ij}^\epsilon$ $(\epsilon=+, -)$ of the maximal compact subalgebra
$u(\infty) \oplus u(\infty)$ of $u(\infty, \infty)$ and of the
noncompact raising and lowering operators $X_{ij}$ and $X_{ij}^*$, respectively
($i,j=1,2,\dots$)
satisfying
\beqa
[E^+_{ij},E^+_{kl}] = \delta_{jk} E^+_{il} - \delta_{il} E^+_{kj}, \hspace{10pt}
[E^-_{ij},E^-_{kl}] = \delta_{jk} E^-_{il} - \delta_{il} E^-_{kj}, \hspace{10pt}
[E^+_{ij},E^-_{kl}] = 0, \hspace{-16pt} \nonumber
\eeqa \vskip-9mm
\beqa
[E^+_{ij},X^*_{kl}] = \delta_{jl} X^*_{ki},\qquad
[E^+_{ij},X_{kl}] = - \delta_{il} X_{kj},\nonumber \\[1mm]
[E^-_{ij},X^*_{kl}] = \delta_{jk} X^*_{il}, \qquad
[E^-_{ij},X_{kl}] = - \delta_{ik} X_{jl}, \nonumber
\eeqa \vskip-9mm
\beqa \label{eq4.1it}
[X_{ij},X_{kl}^*] = \delta_{ik} E^+_{lj} + \delta_{jl} E^-_{ki} \;.
\eeqa
The commuting diagonal elements $E_{ii}^\epsilon$ span a compact
{\it Cartan subalgebra}. The {\it antisymmetric bifield} $A$ gives
rise to an {\it abelian algebra} spanned by the {\it raising
  operators} $Y_{ij}^+ = -Y_{ji}^+$, the {\it lowering operators}
$(Y_{ij}^-)^* = -(Y_{ji}^-)^*$ and the operators $F_{ij}$;
the modes of $A^*$ are hermitian conjugate to those of $A$.
The above $E$'s together with the $F_{ij}$ and their conjugates,
$F_{ij}^*$, give rise to the maximal compact subalgebra $u(2\infty)$
of $so^*(4\infty)$.
The additional nontrivial CR can be restored (applying
when necessary hermitian conjugation) from the following ones:
\beqa\label{eq4.2it}
[E_{ij}^- , F_{kl}] = \podr \delta_{jk} F_{il},\ \ [F_{ij}, E_{kl}^+] = \delta_{jk} F_{il},\ \
[F_{ij}, F_{kl}^* ] = \delta_{jl} E_{ik}^- - \delta_{ik} E_{lj}^+ ;
\nn {}
[X_{ij}, F_{kl}] = \podr \delta_{ik} Y_{jl}^+,\ \ [X_{ij}, F_{kl}^* ] = -\delta_{jl} Y_{ik}^- ,
\nn {}
[Y_{ij}^\epsilon , E_{kl}^\epsilon ] = \podr \delta_{jk} Y_{il}^\epsilon - \delta_{ik}Y_{jl}^\epsilon,\ \
\nn {}
[Y_{ij}^+ , X_{kl}^* ] = \podr \delta_{il} F_{kj} - \delta_{jl} F_{ki},\ \
\nn {}
[Y_{ij}^- , X_{kl}^*] = \podr \delta_{jk} F_{il}^* - \delta_{ik} F_{jl}^*;
\nn {}
[Y_{ij}^\epsilon , (Y_{kl}^\epsilon)^* ] = \podr \delta_{ik} E_{lj}^\epsilon - \delta_{jk} E_{li}^\epsilon
+ \delta_{jl} E_{ki}^\epsilon - \delta_{il} E_{kj}^\epsilon;
\nn {}
[Y_{ij}^+ , F_{kl}^* ] = \podr \delta_{il} X_{kj} - \delta_{jl} X_{ki},\ \
\nn {}
[Y_{ij}^- , F_{kl} ] = \podr \delta_{jk} X_{il} - \delta_{ik} X_{jl}.
\eeqa
We note that the CR (\ref{eq4.1it}) and (\ref{eq4.2it}) do not depend
on the ``central charge'' $2N$ of the inhomogeneous terms in
Eqs. (\ref{eq3.13it}) and (\ref{eq3.14it}) that is absorbed in the
definition of $E_{ii}^\epsilon$ (cf.\ Eq.\ (\ref{eqA.3}) of Appendix
A). The parameter $N$ reappears, however, in the expression for the
{\it conformal Hamiltonian} $H_c$ which involves an infinite sum of
Cartan modes -- and hence only belongs to an appropriate completion of
$u(\infty, \infty)\subset so^*(4\infty)$:
\beq
\label{eq4.3}
H_c = \mathop{\sum}\limits_{i \, = \, 1}^\infty\varepsilon_i (E_{ii}^+ + E_{ii}^- - 2N).
\eeq
Here the energy eigenvalues $\varepsilon_i$ form an increasing
sequence of positive integers (in $D=4$: $\varepsilon_1 = 1, \varepsilon_2 =
\cdots = \varepsilon_5 = 2, \, \varepsilon_6 = \cdots =  \varepsilon_{14} =
3$, etc.).
The {\it charge} $Q$ and the {\it number operator} $C_1^u$
which span the centre of $u(\infty, \infty)$ and of $u(2\infty)$,
respectively, also involve infinite sums of Cartan modes:
\beq
\label{eq4.4}
Q = \mathop{\sum}\limits_{i \, = \, 1}^\infty (E_{ii}^+ - E_{ii}^-), \quad
C_1^u = \mathop{\sum}\limits_{i \, = \, 1}^\infty (E_{ii}^+ + E_{ii}^- - 2N).
\eeq
A priori $N$ is a (positive) real number.
It has been proven in \cite{NST02, NRT07},
however, that in a unitary positive energy realization of any algebra of
bifields generated by local scalar fields of scaling dimension two,
$N$ must be a natural number.

Let us define the {\it vacuum representation} of the bifields $W$ and
$A^{(*)}$ obeying the CR (\ref{eq3.13it}) and (\ref{eq3.14it}) as the
{\it unitary irreducible positive energy representation}
(UIPER) of $so^*(4\infty)$ in which $H_c$ is well
defined and has eigenvalue zero on the ground state
$|vac\rangle$ (the {\it vacuum state}). We are now ready to
state our main result.

\medskip

\noindent {\bf Theorem 4.1} {\it In any UIPER {\rm(}of\/ fixed $N${\rm)} of\/
$so^*(4\infty)$ we have{\rm:}

{\rm(i)} $N$ is a nonnegative integer and all UIPERs of\/ $so^*(4\infty)$ are
realized {\rm(}with multiplicities{\rm)}
in the Fock space $\mathcal{F}_{2N}$ of\/
$2N$ free complex massless scalar fields {\rm(}see Appendix A{\rm)}.

{\rm(ii)} The ground states of equivalent UIPERs of\/ $so^*(4\infty)$ in
$\mathcal{F}_{2N}$ form irreducible representations of the gauge group
$Sp(2N)$. This establishes a one-to-one correspondence between UIPERs
of\/ $so^*(4\infty)$ occurring in the Fock space and the irreducible
representations of\/ $Sp(2N)$.}

\medskip

The {\it proof} parallels that of Theorem 1 in \cite[Sect.~2]{BNRT07}
using the results of Appendix A. We shall only note that each UIPER of
$so^*(4\infty)$ is expressed in terms of the fundamental weights
$\Lambda_\nu$ of $so^*(4n)$ (for large enough $n$, exceeding $N$):
\beq
\label{eq4.5}
\Lambda = \mathop{\sum}\limits_{\nu \, = \, 0}^{2n-1}k_\nu
\Lambda_\nu, \quad k_\nu\leq 0.
\eeq
In particular, the vacuum representation has weight $-2N\Lambda_0$
(see (\ref{eqA.16})). Thus, each UIPER remains irreducible when
restricted to some $so^*(4n)$, so that we are effectively dealing with
representations of finite dimensional Lie algebras. We also note that
the bifield $W$ has a vanishing vacuum expectation value in view of
(\ref{eqA.15}), in accord with its definition as a sum of twist two
local fields.

The outcome of Theorem 4.1
and of Theorems 1 and 3 of \cite{BNRT07} was expected in
view of the abstract results of the
Doplicher--Haag--Roberts theory of superselection
sectors \cite{H92, DR90, BDLR92}. However,
considerable technical difficulties are
encountered in relating the extension theory of bifields with the
representations of the corresponding nets. Our study provides an independent
derivation of DHR-type results in the field theoretic framework, advancing at
the same time the program of classifying globally conformal invariant quantum
field theories in four dimensions.

\section*{Acknowledgments}

The results of the present paper were reported at three conferences
during the summer of 2007: ``LT7 -- Lie Theory and Its Applications in
Physics'' (Varna, June 18--23); ``Infinite Dimensional Algebras and
Quantum Integrable Systems'' (Faro, July 23--27); ``SQS'07 --
Supersymmetries and Quantum Symmetries'' (Dubna, July 30 -- August
4). I.T. thanks the organizers of all three events for hospitality and
support. N.M.N.\ and K.-H.R. also acknowledge the invitation to the
Varna meeting. B.B.\ was partially supported by NSF grant
DMS-0701011. N.M.N.\ and I.T.\ were supported in part by the Research
Training Network of the European Commission under contract
MRTN-CT-2004-00514 and by the Bulgarian National Council for Scientific
Research under contract PH-1406. K.-H.R.\ thanks the
Alexander-von-Humboldt foundation for financial support.

\appendix

\renewcommand{\thesection}{Appendix~\Alph{section}.}
\section{Fock space realization of the Lie algebra $so^* (4n)$ (for $n \to
  \infty$)}\label{Ap.A}
\setcntrs
\renewcommand{\thesection}{\Alph{section}}
\renewcommand{\theequation}{\Alph{section}.\arabic{equation}}

For the higher dimensional vertex algebra formalism (and the associated
complex variable realization of compactified Minkowski space) used in this
Appendix, see \cite{BN06} and references therein (for a summary, see
Appendix~A to \cite{BNRT07}).
We write the pair of vectors of complex fields (\ref{eq3.12it}) as
\beq
\psi (\z) = a (\z) + b^* (\z) \,, \quad
\psi^*(\z) = a^* (\z) + b (\z) \,,
\eeq
where $\psi = (\VEC{\psi}_\alpha :\alpha = 1, 2) = (\psi_\alpha^p :\alpha = 1,
2,\; p=1,\ldots, N)$ and likewise for $a,b$. Their mode decomposition in
the compact picture is:
\beqa\label{eqA.1}
\VEC{a}_{\alpha}(\z) \, = \podr \mathop{\sum}\limits_{\ell \, = \, 0}^{\infty}
\frac{1}{\sqrt{\ell+1}} \, \mathop{\sum}\limits_{\mu \, = \, 1}^{(\ell+1)^2}
\frac{\VEC{a}_{\alpha n}}{\bigl(\z^2\bigr)^{\ell+1}} \ h_{\ell,\mu} (\z)
\, , \quad
\nn
\VEC{b}^*_{\beta}(\z) \, = \podr \mathop{\sum}\limits_{\ell \, = \, 0}^{\infty}
\frac{1}{\sqrt{\ell+1}} \, \mathop{\sum}\limits_{\mu \, = \, 1}^{(\ell+1)^2}
\VEC{b}_{\beta n} \, h_{\ell,\mu} (\z) \, , \quad
\eeqa
where $\{h_{\ell,\mu} (\z): \mu=1,\dots,(\ell+1)^2\}$ is a basis of
homogeneous harmonic polynomials of degree $\ell$ in the $4$-vector $\z$,
diagonalizing the conformal one-particle energy,
$n=n(\ell,\mu)$ ($=1,2,\dots$) is an enumeration, and
$a_n^{(*)}$, $b_n^{(*)}$ obey the
canonical commutation relations (we only list the nontrivial ones):
\beq\label{eqA.2}
\bigl[a_{\alpha m}^p,a^{q*}_{\beta n}\bigr]
\, = \,
\delta_{\alpha\beta} \,
\delta_{mn} \, \delta^{pq}
\, = \,
\bigl[b_{\alpha m}^p,b^{q*}_{\beta n}\bigr]  \,.
\eeq
The corresponding modes of the bifield $W$ from \eqref{eq3.11it}
(i.e., the $u (\infty,\infty)$
generators) are split into two groups.
First, we have the compact ($u(\infty) \oplus u(\infty)$) generators:
\beq\label{eqA.3}
E_{ij}^+ \, = \, \frac{1}{2}\, \bigl[a_i^*,a_j\bigr]_+ \, = \,
a_i^* a_j + N \delta_{ij} \, , \quad
E_{ij}^- \, = \, \frac{1}{2}\, \bigl[b_i^*,b_j\bigr]_+ \, = \,
b_i^* b_j + N \delta_{ij} \,,
\eeq
where $a^*_i a_j$, etc., stand for the inner products
\beq\label{eqA.4}
a_i^* a_j \, = \, \mathop{\sum}\limits_{\alpha \, = \, 1}^2 \,
\VEC{a}^*_{\alpha i} \spr \VEC{a}_{\alpha j}
\, = \,
\mathop{\sum}\limits_{\alpha \, = \, 1}^2
\mathop{\sum}\limits_{p \, = \, 1}^N
a_{\alpha i}^{p *} \, a_{\alpha j}^{p} \,.
\eeq
Second, we have the energy decreasing ($X_{ij}$) and energy increasing
($X_{ij}^*$) operators:
\beq\label{eqA.5}
X_{ij} \, = \, b_i a_j \,
(\, \equiv \, \mathop{\sum}\limits_{\alpha \, = \, 1}^2 \VEC{b}_{\alpha i} \spr \VEC{a}_{\alpha j})
\, , \quad
X_{ij}^* \, = \, b_i^* a_j^* \,
\,.
\eeq
The modes of the skewsymmetric bifield $A$ from \eqref{eq3.11it}
and its conjugate also
include a compact part ($F_{ij}^{(*)}$) and a noncompact one
($Y^{\pm (*)}_{ij}$):
\beqa\label{eqA.6}
F_{ij} \, = \podr \VEC{b}_{1i}^* \spr \VEC{a}_{2j} - \VEC{b}_{2i}^*
\spr \VEC{a}_{1j} \,; \\
Y_{ij}^+ \, = \podr \VEC{a}_{1i}^* \spr \VEC{a}_{2j} - \VEC{a}_{2i}^*
\spr \VEC{a}_{1j} \, (\, = - Y_{ji}^+) \,,
\nn \label{eqA.7}
Y_{ij}^-\, = \podr \VEC{b}_{1i}^* \spr \VEC{b}_{2j} - \VEC{b}_{2i}^*
\spr \VEC{b}_{1j} \, (\, = - Y_{ji}^-) \,
\eeqa
and their conjugates.

We shall now present the subalgebras obtained by restricting the modes
to the $n$ lowest one-particle energies. Because we wish to treat
positive-energy representations as highest-weight representations, it is
convenient to assign positive roots to energy lowering operators.
According to the ordering of energies $\varepsilon_1^- =
\varepsilon_1^+ \leqslant \varepsilon_2^- = \varepsilon_2^+ \leqslant
\cdots$ (dealing with degeneracies as in \cite{BNRT07}) we choose
the (ordered set of) simple roots and raising operators (= energy
lowering operators) as
\beq\label{eqA.8}
\begin{array}{lll}
\alpha_0  = - e_1 - e_2
 , &\hspace{8pt} H_0  =  - E_{11}^- - E_{11}^+
 , &\hspace{8pt}  X_{11} \, (\equiv E_0)
 ,
\\
\alpha_1  = e_1 - e_2
 , &\hspace{8pt} H_1  =  E_{11}^- - E_{11}^+
 , &\hspace{8pt} F_{11} \, (\equiv E_1)
 ,
\\
\alpha_2  = e_2 - e_3
 , &\hspace{8pt} H_2  =  E_{11}^+ - E_{22}^-
 , &\hspace{8pt} F_{12}^* \, (\equiv E_2)
 ,
\\
\cdots &\hspace{8pt} \cdots&\hspace{8pt} \cdots
\\
\alpha_{2n-2}  =  e_{2n-2} - e_{2n-1}
 , &\hspace{8pt} H_{2n-2}  =  E_{n-1 n-1}^+ - E_{nn}^-
 , &\hspace{8pt} F_{n-1n}^* \, (\equiv E_{2n-2})
 ,
\\
\alpha_{2n-1}  = e_{2n-1} - e_{2n}
 , &\hspace{8pt} H_{2n-1}  =  E_{nn}^- - E_{nn}^+
 , &\hspace{8pt} F_{nn} \, (\equiv E_{2n-1}) \,.
\end{array}
\eeq
Here
the names $H_\nu$ and $E_\nu$ of the generators comply with the
standard Chevalley--Serre notation; the vectors
$\{e_{s}\}$ form an orthonormal basis so that the scalar products
$(\alpha_i|\alpha_j)$ reproduce the Cartan matrix of $so^*(4n)$:
\beqa\label{eqA.9}
(\alpha_i|\alpha_i) \, = \podr 2\,,
 \qquad
(\alpha_0|\alpha_1) \, = \, 0
\, , \quad
\nn
(\alpha_0|\alpha_2) \, = \podr (\alpha_1|\alpha_2)
\, = \, -1 \, = \, (\alpha_i|\alpha_{i+1}) \,
\eeqa
for $i=2,\dots,2n-2$. The positive roots (corresponding to the raising
operators) are $e_i-e_j$ and $-e_i-e_j$ ($1\leq i < j \leq 2n$).

The sum $t$ of the vectors $e_{s}$ is a root vector,
the corresponding Cartan element $H_t$
generating the center of
the maximal compact Lie subalgebra $u(2n)$ of $so^* (4n)$:
\beq\label{eqA.10}
t \, = \, \mathop{\sum}\limits_{s \, = \, 1}^{2n} \, e_s
\, , \qquad
H_t \, = \, \mathop{\sum}\limits_{i \, = \, 1}^{n} \, \bigl( E_{ii}^+ + E_{ii}^- \bigr) \,.
\eeq
The $so^* (4n)$ fundamental weights $\Lambda_{\nu}$ and the half sum
$\delta$ of positive roots of $so^*(4n)$
are given by
\beqa\label{eqA.11}
\Lambda_0 \, = \podr - \frac{t}{2} \, , \quad
\Lambda_1 \, = \, e_1 - \frac{t}{2} \,,
\nn
\Lambda_j \, = \podr \mathop{\sum}_{s \, = \, 1}^j e_s -t
\, = \, - \sum_{s=j+1}^{2n} e_s
\quad (j \, = \, 2,\dots,2n-1\,), \qquad
\\ \label{eqA.12}
\delta \, = \podr \mathop{\sum}_{\nu \, = \, 0}^{2n-1} \Lambda_{\nu}
\, = \, -\sum_{s=1}^{2n}(s-1)e_s \,=\,\rho - \bigl(n-\frac{1}{2}\bigr) \, t
\eeqa
where $\rho$ is the half sum of positive roots of $su(2n)$:
\beq\label{eqA.13}
\rho \, = \, \mathop{\sum}_{s \, = \, 1}^n \Bigl(n-s+\frac{1}{2}\Bigr) \bigl(e_s - e_{2n+1-s}\bigr) \,.
\eeq
Note that $(t|\alpha)$ $= 0 = (t|\rho)$
for $\alpha$ a root of $su(2n)$; observe
that $\rho,\delta$ and the Casimir invariants below are only defined
for finite $n$. The second order Casimir operators of
$so^* (4n) \supset u(2n) \supset su(2n)$ are related by
\beqa\label{eqA.14}
C_2^{u(2n)}
\, = \podr C^{su (2n)}_2 + \frac{H_t^2}{2n}
\, = \, C^{so^* (4n)}_2 + 2 \, \mathop{\sum}_{j \, = \, 1}^n X_{ij}^* X_{ij}
\nn
\podr
+ \, 2 \, \mathop{\sum}_{1 \, \leqslant \, i \, < \, j \, \leqslant \, n}
\bigl( Y_{ij}^{+ *} Y_{ij}^+ + Y_{ij}^{- *} Y_{ij}^- \bigr)
+ \bigl(2n-1\bigr) H_t \,.
\eeqa

The \textit{vacuum} $\vac$ is defined as a basis vector in
a $1$-dimensional space satisfying the relations $\vec a_{\alpha i}
\vac = 0 = \vec b_{\alpha i} \vac$,
or equivalently
\beq\label{eqA.15}
X_{ij} \vac \, = \, Y_{ij}^{\pm} \vac \, = \, 0 \, = \, F_{ij}^{(*)} \vac \, , \quad
E_{ij}^{\pm} \vac \, = \, N \delta_{ij} \, \vac \,.
\eeq
It follows that it can be identified with the highest weight vector of
a unitary irreducible representation of $so^* (4n)$ (for any $n>1$) of
weight $-2N\Lambda_0$:
\beq\label{eqA.16}
\vac \, = \, |\!-2N\Lambda_0\rangle
\, , \quad C_2^{so^* (4n)} \bigl(-2N\Lambda_0\bigr) \, = \, 2nN\bigl(N+1-2n\bigr) \,.
\eeq

As anticipated by the ordering (\ref{eqA.8}) of roots
(and of Cartan and raising operators),
it is convenient to relabel the oscillators by setting
\beq\label{eqA.17}
\VEC{A}_{2i-1} = \, - \VEC{b}_{2,i}
\, , \quad
\VEC{A}_{2i} = \, \VEC{a}_{1,i}
\, , \quad
\VEC{B}_{2i-1} = \, \VEC{b}_{1,i}
\, , \quad
\VEC{B}_{2i} = \, \VEC{a}_{2,i}
\, , \quad
i \, = \, 1,\dots,n\,.
\eeq
Then the generators of $so^* (4n)$ can be rewritten as
\beq\begin{array}{rll}E^\pm,F,F^*&\to& E_{kl} = \vec A\,_k^*\cdot \vec
  A_l + \vec B\,_k^*\cdot \vec B_l + \delta_{kl} N, \\ X,Y^\pm &\to&
  Y_{kl} = \vec A_k \cdot \vec B_l - \vec B_k \cdot \vec A_l
\qquad\qquad (k,l=1,\dots 2n).\end{array}\eeq
We refrain
from displaying the commutation relations of $so^*(4n)$ again, which are most
easily (and more compactly than (\ref{eq4.2it})) read off this representation.

Our aim is to classify the UIPERs of $so^*(4n)$
with ground states $\vert h\rangle$
with Cartan eigenvalues
\beq E_{kk} \vert h\rangle = h_k \vert h\rangle.\eeq
We omit the details of the argument, which is in perfect analogy with
\cite{BNRT07}, indicating only the three main steps.

1. Unitarity of the submodule $U(h)$ obtained by acting with the
generators of the maximal compact Lie subalgebra $u(2n)$ on the ground
state implies that
\beq h_1 \geq h_2 \geq h_3 \geq \cdots \eeq
is an integer-spaced non-increasing sequence, stabilizing at some value
$h_\infty$, and $h_\infty = 2N/2 = N$ in order to have a finite Hamiltonian.
The finiteness of the operators $Q$ and $C_1^u$ on all states of
  finite energy is then
  automatically guaranteed.

2. We choose $n$ large enough so that $h_{2n} = h_\infty =N$. Let $\Young$ be
the Young tableau of $su(2n)$ with rows of length $m_k=h_k-h_\infty$.

The noncompact generators $Y^*_{kl}$ with negative roots
transform like the antisymmetric rank 2 representation of $u(2n)$.
Hence, the linear span of $Y^*\,U(h)$ decomposes into
irreducible representations of $u(2n)$ whose Young diagrams are
obtained by adding two boxes in different rows to $\Young$. Their highest
weights $\lambda$ are of the form $h + e_k + e_l$ where $k\neq l$.

In each of these states, the above Casimir operators can be computed.
Since the difference $C_2^{u(2n)}-C_2^{so^*(4n)}$ is a positive
operator, the difference of eigenvalues must be nonnegative. This
yields the necessary bounds
\beq (\lambda + \delta,\lambda + \delta) - (h+\delta,h+\delta)  \geq 0 \eeq
for all $\lambda=h+e_k+e_l$. The strongest bound occurs
when $k$ and $l$ are chosen maximal, i.e., $k=r+1$ and $l=r+2$ when
$r+1$ is the smallest index such that $h_{r+1}=h_\infty$ (i.e., $r$ is
the number of the rows of the Young diagram $\Young$). Evaluating
the bound, yields the condition
\beq r \leq N. \eeq

3. The Young diagrams admitted by this condition are precisely those of
the unitary tensor representations of $U(N)$. It remains to establish
the relation between these and the unitary representations of the
gauge group $Sp(2N)$ of the field algebra (\ref{eq3.13it}) (which
contains $U(N)$), and to verify that each of these is realized on the Fock space
of $2N$ complex massless free fields $\psi_{\alpha}^p$ ($\alpha=1,2$,
$p=1,\dots,N$).

By the above relabeling of the oscillators, the infinitesimal
generators of $sp(2N)$ become
\beqa
E^{pq} = E^{qp*} &=& \sum_{k=1}^{2\infty}(A_k^{p*}A_k^q - B_k^{q*}B_k^p) \,,
\\
X^{pq} = X^{qp} &=& \sum_{k=1}^{2\infty}(A_k^{p*}B_k^q +
A_k^{q*}B_k^p) \,, \qquad p,q=1,\dots,N\,.
\eeqa
The $E^{pq}$ are the generators of
$U(N)\subset Sp(2N)$, and $A^*$ (the creation operators for
$\vec\psi_1^*$ and for $\vec\psi_2$) transform in the vector
representation of $U(N)$, while $B^*$ (the creation operators for
$\vec\psi_2^*$ and for $\vec\psi_1$) transform in the conjugate
representation. In other words, one may assign the weights $e^p$ to
$A^{p*}$ and $-e^p$ to $B^{p*}$, so that $E^{pq}$ correspond to the
roots $e^p - e^q$ and $X^{pq}$ to
$-e^p-e^q$. The simple roots are $e^p-e^{p+1}$ (corresponding to
$SU(N)\subset Sp(2N)$) and $2e^N$.

Now let $(h_1,h_2,\dots,h_N,h_{N+1} = \cdots = h_n = N)$ be the Cartan
weights of a positive-energy representation of $so^*(4n)\subset so^*(4\infty)$.
Let $\Young$ be the Young diagram of $U(N)$ with rows of
length $m_k=h_k-N$, and $r_l$ the heights of its columns.

Define in the Fock space of the complex free fields $\vec\psi_1^{(*)}$ and
$\vec\psi_2^{(*)}$ the vector
\beq \vert h \rangle_F = \left(\prod_{l=1}^{m_1} A^{*\wedge
    r_l}\right) \, \vac \eeq
where $A^{*\wedge r} = \det\Big(A^{p*}_k\Big)_{k=1,\dots r}^{p=1,\dots r}$.
Then $\vert h \rangle_F$ is a highest weight vector for
$so^*(4\infty)$ with the proper Cartan eigenvalues $h_k$ of $E_{kk}$.
It is a component of a $U(N)$ tensor in the representation given
by $\Young$. This tensor extends, by the action of the generators $X^{pq}$,
$X^{pq*}$, to a $Sp(2N)$ tensor. (The generators $X^{pq*}$ will swap
some of the $A$-excitations into $B$-excitations.)

As a representation of $u(N)$, this representation has highest weight
$w = m_1e^1 + \cdots + m_Ne^N$. We decompose this into the fundamental
weights of $sp(2N)$. These are
determined by the property that $(\Lambda^l,\alpha_k) = \delta^l_{k}$
  where $\alpha_k$ are the simple roots, giving
$\Lambda^l=e^1+\cdots +e^l$ ($l=1,\dots N-1$) and
$\Lambda^N = \frac 12 \sum_{p=1}^Ne^p$. Then
\beq w = n_1 \Lambda^1 + \cdots + n_{N}\Lambda^{N}\eeq
with $n_l=m_l-m_{l+1}$ ($l<N$), and $n_N = 2m_N$. We
therefore obtain all those representations of $sp(2N)$ for which
$n_N$ is even.

Representations with half-integral weights ($n_N$ odd)
integrate to representations of a two-fold covering of $Sp(2N)$,
because the $U(1)$ subgroups $\exp itE^{pp}$ integrate to $-1$ as
$t=2\pi$. Thus, we obtain the desired duality
result: All irreducible positive-energy representations of
$su^*(4\infty)$ are realized on the Fock space, and their multiplicity
spaces are representation spaces of all irreducible unitary true
representations of the gauge group $Sp(2N)$.

\addtocontents{toc}{\contentsline {section}{References}{\arabic{page}}}

\end{document}